\documentclass[usenatbib]{basi}
\usepackage[british]{babel}
\usepackage[varg]{txfonts}

\usepackage{epstopdf}

\begin{document}
\title[Searches for radio transients]{Searches for radio transients}
\author[N.~D.~R.~Bhat]%
       {N.~D.~R.~Bhat$^{1,2}$\thanks{email: \texttt{rbhat@astro.swin.edu.au}} \\
        $^1$Centre for Astrophysics and Supercomputing, Swinburne University of Technology, Hawthorn,
          \\ Victoria 3122, Australia\\
        $^2$Australian Research Council Centre of Excellence for All-Sky Astrophysics (CAASTRO)
        }

\pubyear{2011}
\volume{39}
\pagerange{\pageref{firstpage}--\pageref{lastpage}}

\date{Received 2011 September 18; accepted 2011 October 03}

\maketitle
\label{firstpage}

\begin{abstract}
Exploration of the transient Universe is an exciting and fast-emerging area within radio astronomy.
Known transient phenomena range in time scales from sub-nanoseconds to years or longer, thus
spanning a huge range in time domain and hinting a rich diversity in their underlying physical processes.
Transient phenomena are likely locations of explosive or dynamic events and they offer tremendous
potential to uncover new physics and astrophysics.
A number of upcoming next-generation radio facilities and recent advances in computing and
instrumentation have provided a much needed impetus for this field which has remained a relatively
uncharted territory for the past several decades.
In this paper we focus mainly on the class of  phenomena that occur on very short time scales
(i.e.\ from $\sim$milliseconds to $\sim$nanoseconds), known as {\it fast transients}, the detections
of which involve considerable signal processing and data management challenges, given the high time
and frequency resolutions required in their explorations, the role of propagation effects to be
considered and a multitude of deleterious effects due to radio frequency interference.
We will describe the techniques, strategies and challenges involved in their detections and review the
world-wide efforts currently under way, both through scientific discoveries enabled by the ongoing
large-scale surveys at Parkes and Arecibo, as well as technical developments involving the
exploratory use of multi-element array instruments such as VLBA and GMRT.
Such developments will undoubtedly provide valuable inputs as next-generation arrays such as LOFAR
and ASKAP are designed and commissioned. With their wider fields of view and higher sensitivities,
these instruments, and eventually the SKA, hold great potential to revolutionise this relatively nascent
field, thereby opening up exciting new science avenues in astrophysics.
\end{abstract}



\section{Introduction}\label{s:intro}

Exploring the transient Universe has been a major astrophysical frontier over the past several decades.
Transient phenomena are thought to be likely locations of explosive or dynamic events, thereby providing
enormous potential to uncover a wide range of new physics and astrophysics (e.g.\ Cordes et al.\ 2004).
The transient high-energy sky (X- and $\gamma$-ray wavelengths) is perhaps the most well explored within
the electromagnetic spectrum, where a number of space-based wide field-of-view instruments with all-sky
monitors onboard routinely scan the sky for exotic events.
Such instruments have led to great successes in finding sources such as gamma-ray bursts and accreting sources.
At optical wavelengths, searches for phenomena such as gravitational lensing events and Type Ia supernovae have
led to the development of high throughput instruments such as PanSTARRs and LSST. While radio astronomy has an
impressive record of achieving high time resolution, the lack of suitable instruments and the tremendous signal
processing overheads, have greatly limited our ability to explore the transient sky at radio wavelengths.
The transient radio Universe is thus by and large a largely uncharted territory.

The exploration of the transient radio sky has primarily been hampered by the lack of instruments with
wide-field capabilities at radio wavelengths.
Thanks to a number of upcoming next-generation facilities, this inherent technological limitation will
soon be overcome, thereby opening up an exciting era in radio astronomy. A number of new radio facilities are
either in their commissioning phase or under construction, many of which will offer wide field-of-view capabilities.
These include low-frequency arrays such as Low Frequency Array (LOFAR) and the Murchison Widefield Array (MWA),
as well as the Square Kilometer Array (SKA) pathfinder instruments, viz.\ the Australian SKA Pathfinder (ASKAP)
in Western Australia and MeerKAT in South Africa (R\"ottgering et al.\ 2006; Johnston et al.\ 2007; Lonsdale et al.\ 2009).
Exploring the transient radio sky is also a key science driver for the SKA
(e.g.\ Wilkinson et al.\ 2004, Cordes et al.\ 2004; Cordes 2009).
The availability of such next-generation arrays, together with appropriately designed instrumentation and
suitable data archival and processing strategies, can potentially revolutionise our knowledge of the transient
radio sky  in the coming decades.

Transient phenomena are known to range in time scales from sub-nano seconds and longer. For example,
giant radio pulses from the Crab pulsar are unresolved down to 0.4 ns (Hankins \& Eilek 2007), implying
apparent brightness temperatures as much as $10^{42}$~K. Even at traditional radio astronomy frequencies
($\sim 1{-}2$~GHz) where the achievable time resolution is limited by interstellar propagation effects such
as multi-path scattering, the brightness temperatures are known to reach as much as $10^{35}$~K
(Bhat, Tingay \& Knight 2008). Phenomena such as supernovae light curves and gamma-ray-burst after glows
have rather long time scales of the order of years. Thus the time scales to explore can span a huge range
of almost 20 orders of magnitude.

A distinction can be made between ``slow'' and ``fast'' transients within the context of radio astronomy.
A popular definition is given by Cordes (2009). In brief, for radio telescopes with conventional fields of view
(e.g.\ paraboloids with single-pixel feeds), slow transients are those that can be sampled in a raster-scan
survey, because they stay on for at least as long as it takes to scan the relevant sky region. Fast transients,
conversely, are those that would be missed in the time it takes to scan the sky. In terms of the technical
requirements, studies of slow transients require imaging on a wide range of time integrations, e.g.
from snapshot to daily, while fast transients require time-domain signal processing of data sampled at high time
and frequency resolutions.  The computational needs can be substantial for the exploration
of short-duration phenomena where time scales are in the sub-second regime.  Furthermore, as
is well known, impulsive radio frequency interference (RFI) can be a major impediment in making credible detections
(e.g.\ Bhat et al.\ 2005; Burke-Spolaor et al.\ 2011a). Not only that such interference signals can potentially
mimic one or more signatures of real signals, their frequent occurrence may also impact the achievable sensitivity,
making weaker signals difficult to detect.

Fortuitously, the computational side of radio transient research is fast becoming less of a challenge,
with several impressive advances being made in affordable and efficient supercomputing. For example,
the use of Graphics Processing Units (GPUs) for astronomy computing is rapidly gaining popularity, as their
suitability is vividly demonstrated for a number of applications
(e.g.\ Wayth, Greenhill \& Biggs 2009; Barsdell, Barnes \& Fluke 2010; Margo et al.\ 2011). RFI still poses a formidable
challenge however, given its tendency to manifest as numerous false positives even with sophisticated search
algorithms and processing strategies.

Transients on very short time scales are often linked to coherent radiation and, frequently, to sources in
extreme matter states by making a simple light-travel size argument.
Taking pulsar phenomena as examples, the implied brightness temperatures can be very high; e.g.
$10^{20}$ to $10^{42}$~K on timescales of milliseconds to nanoseconds.  Such short-duration transients are naturally
affected by plasma propagation effects such as dispersion, multi-path scattering and scintillation by
the intervening media (e.g.\ interplanetary, interstellar and/or intergalactic media). For these same reasons,
they may also serve as excellent probes of the intervening media.

Much of the current fast transient exploration makes use of large single-dish instruments such as the Parkes
and Arecibo telescopes (e.g.\ Deneva et al 2009; Burke-Spolaor et al 2011b), which at best, can provide limited
resilience against RFI via useful cross checks  possible between simultaneous multiple data streams of the
multi-pixel receivers. Array instruments can potentially provide a much higher resilience to RFI, provided
their long baseline and interferometric advantages can be exploited for effective identification and excision
of RFI-generated transient events. Efforts have already begun in this direction through the exploratory use
of existing instruments such as VLBA and GMRT. This paper will describe the basic methodologies and search
algorithms used in transient detections and present an overview of the past and ongoing efforts aimed at
transient searches, as well as those proposed with upcoming instruments including the SKA.


\section{Discovery potential of radio transients}\label{s:phase}

Besides pulsar radio emission, which is known to occur on a variety of time scales, ranging from milliseconds
(i.e.\ sub-pulses), to microseconds (i.e.\ micro-structure) and even to nanoseconds (i.e.\ giant pulses), and
solar and stellar flares, most known radio transients have been found from follow-up observations of targets
selected from surveys at other wavelengths and high energies.
Radio afterglows from gamma ray bursts (GRBs) are well known and are discussed in another paper in this issue
(Chandra \& Frail 2011).
The detection of periodic pulsations from magnetars such as XTE J1810$-$197 and 1E 1547.0$-$5408 are other
examples (Camilo et al.\ 2006, 2007). An interesting exception is the discovery of the third magnetar known
to emit at radio wavelengths, PSR J1622$-$4950, which was first discovered via its radio emission
(Levin et al.\ 2010) and subsequently followed up at other wavelengths such as X-rays.
Other interesting discoveries include transient sources in the Galatic-center (GC) direction
(Hyman et al.\ 2005; Bower et al.\ 2007) found through VLA imaging observations of the GC, and a specific
class of neutron stars, called rotating radio transients, that were found through large pulsar surveys
using the Parkes and Arecibo telescopes (McLaughlin et al.\ 2006; Keane \& McLaughlin 2011). The detection of
a powerful millisecond burst by Lorimer et al.\ (2007) also  generated much curiosity and stimulated searches
for more such bursts by other radio telescopes around the world (e.g.\ Siemion et al.\ 2011). A close scrutiny
of the detection of more such events (Burke-Spolaor et al.\ 2011a) however suggests that it is unlikely to
be of astrophysical origin.


\begin{figure}
\centerline{\includegraphics[width=11cm]{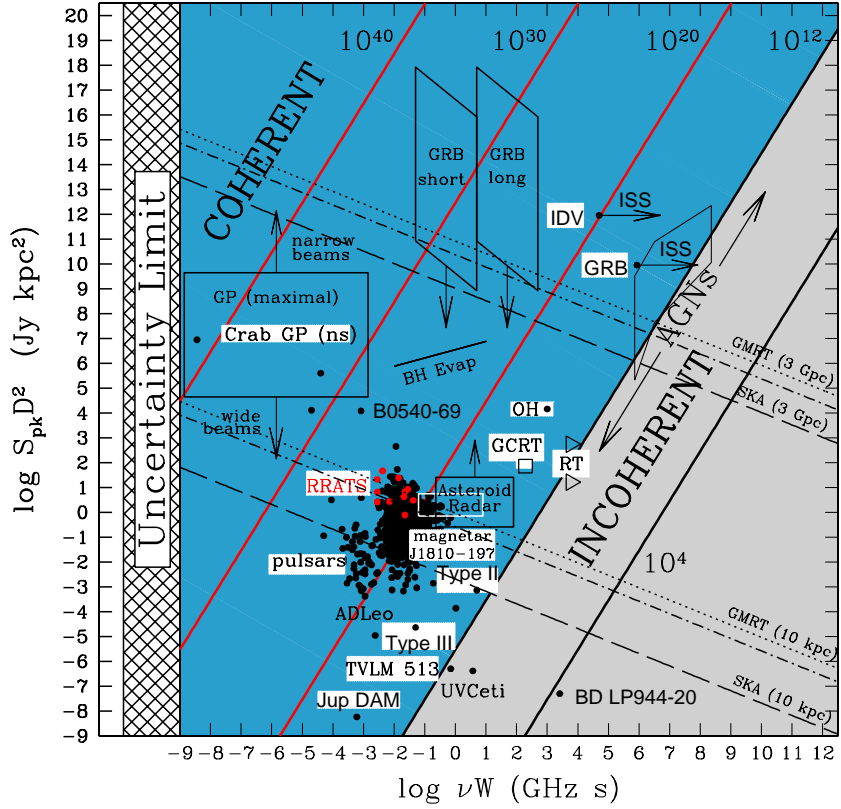}}
\caption{
Time-luminosity phase space for known radio transients from Cordes (2009); a log--log plot of the
product of peak flux  density $S_{\rm pk}$ in Jy and the square of the distance $D$ in kpc vs.\ the product of
frequency $\nu$ in GHz  and pulse width $W$ in s. The uncertainty limit on the left indicates that
$\nu W > 1$ as follows from the uncertainty principle. Lines of constant brightness temperature
$T_{\rm b} = S D^2 / 2 k (\nu W)^2 $ are shown, where $k$ is Boltzmann's constant. Points are shown
for the nano-giant pulses detected from the Crab, giant pulses detected from the Crab pulsar
and a few millisecond pulsars, and single pulses from other pulsars. Points are shown for Jovian
and solar bursts, flares from stars, brown dwarfs, OH masers, and AGNs. The regions labeled
coherent  and incoherent are separated by the canonical $10^{12}$~K limit from the inverse Compton
effect that is relevant to incoherent synchrotron sources.
The growing number of recent discoveries of transients illustrates the fact that empty regions of the
$\nu W{-}S_{\rm pk} D^2$ plane may be populated with sources not yet discovered.
The figure also includes hypothetical transient sources and detection curves; e.g.\ maximal giantpulse
emission from pulsars, prompt radio emission from GRBs, bursts from evaporating black holes, and
radar signals used to track potentially impacting asteroids and comets in exosystems. Long-dashed lines indicate
the detection threshold for the full SKA for sources at distances of 10 kpc and 3 Gpc. Dotted and dot-dashed lines
correspond to the current and future GMRT at 1.2~GHz (i.e.\ bandwidths of 32 MHz and 400 MHz, respectively).
At a given $\nu W$, a source must have luminosity above the line to be detectable. The curves assume optimal
detection (matched filtering).
}
\label{fig:phase}
\end{figure}

While the searches and follow-up studies with existing radio telescopes (i.e.\ limited fields of view) have
led to several impressive results, a closer look at the parameter space for transients suggests they are
nothing more than a tiny subset of what is actually there to uncover or explore.
This is elegantly summarized in the phase space plot (see Figure~\ref{fig:phase}) from Cordes (2009). Various
known transient phenomena or sources are plotted in the time-luminosity space, along with some hypothesized sources.
%
%
Lines of constant brightness temperature ($T_{\rm b}$) are calculated assuming that $W$ is the light travel time across the source,
\begin{equation}
 T_{\rm b} = { S_{\rm pk} \over 2 \, k } \left( { D \over \nu \, W } \right)^2 = 10^{20.5} \, K \, S_{\rm mJy} \, \left( { D_{\rm kpc} \over \nu _{\rm GHz}
\, W_{\rm ms} } \right)^2
\end{equation}
where $S_{\rm mJy}$ is the peak flux density (mJy) at frequency $\nu$ (GHz) and $D_{\rm kpc}$ is the distance (kpc). The
above expression ignores relativistic compression, and consequently for some sources, $W$ can be much smaller than
the light travel time. This would readily imply
exceedingly high equivalent brightness temperatures for giant pulses, reaching as high as $10^{42}$~K for
giant pulses from the Crab pulsar
and $10^{39}$~K for those from the millisecond pulsar B1937+21. Even at typical observing frequencies of 0.4--1 GHz
where pulse broadening from multipath scattering limits the achievable time resolution, $T_{\rm b}$ is known to reach
values as much as $10^{35}$~K (Bhat et al.\ 2008).

In summary, with only a tiny part of the potentially available parameter space explored thus far, the
discovery potential is enormous in this new frontier of astronomy, thereby making a compelling case to conduct
comprehensive sky surveys for radio transients. This recognition has prompted making important changes in the
survey strategies, with most ongoing large surveys for pulsars routinely processing the data for transient
emission on the order of milliseconds or longer. In Section~\ref{s:radio} we will summarise these efforts and
the initial discoveries, along with exploratory surveys that are being launched with array type instruments such
as VLBA and GMRT, and the proposed large-scale surveys with next generation radio arrays such as LOFAR and ASKAP.


\section{Detection techniques, strategies and challenges}\label{s:detect}

We now discuss the detection techniques and search algorithms for fast transient searches. While the basic steps in searching and detection are quite similar, the strategies for identifying prospective candidates and efficiently discriminating them against spurious events of RFI or instrumental origins may significantly differ between different instruments.  We briefly describe the detection sensitivity and parameter space for transient searches as well as the role of propagation effects, in particular at low frequencies. We will then describe the search algorithms as implemented in typical processing chains and highlight some major challenges inherent in fast transient searches.

\subsection{Basic considerations}

\subsubsection{Detection sensitivity}

Sensitivity is an important consideration in the searches for short-duration transients as it defines the maximum distance to which a signal of a given strength is sensitive to ($D_{\rm max}$) and hence the search volume sampled. A transient signal is detectable if its peak flux density
($S_{\rm pk}$) exceeds some minimum flux density as determined by the radiometer equation
\begin{equation}
S_{\rm pk,min} \, =  \, K \, { \beta \, (T_{\rm rec} + T_{\rm sky}) \over G \, \sqrt{ \Delta \nu \, N_{\rm pol} \, W_{\rm p} } }
\end{equation}
where $T_{\rm rec}$ and $T_{\rm sky}$ are the receiver and sky temperatures, respectively (the system temperature, $T_{\rm sys} \approx T_{\rm rec} + T_{\rm sky}$ for most instruments), $G$ is the gain (in $\rm K\,Jy^{-1}$), $\Delta \nu$ is the recording bandwidth and $N_{\rm pol}$ is the number of polarisations; $W_{\rm p}$ is the effective matched filter width employed in transient searching; the factor $\beta$ denotes the loss in signal-to-noise ratio (S/N) due to signal digitization, and the factor $K$ is the detection threshold in units of rms flux density ($\sigma$). As $D_{\rm max} \propto ({S_{\rm pk,min}})^{-1/2}$, for
a given system, the only means of improving the sensitivity for fast transient signal detection is by increasing the bandwidth $\Delta \nu$ over which  observations are made.


For an array comprising of $N$ elements, the net gain $G$ depends on the incoherent vs.\ coherent combination of telescope power, and will thus scale as $\sqrt{N} \, G_{\rm ant} $ vs.\ $N \, G_{\rm ant}$, where $G_{\rm ant}$ is the gain of a single element.
For example, the achievable sensitivity for a GMRT survey at 610|MHz (for $N=30$ and $\Delta \nu=16$~MHz) may range from 1.5~Jy for $W_{\rm p}=1$~ms to $\sim 0.1$~Jy for $W_{\rm p}=100$~ms, assuming a $5\sigma$ threshold.
$T_{\rm sky}$ is a strong function of the observing frequency ($\nu$), with a nominal scaling $\sim \nu^{-2.6}$, and can be expected to vary between different instruments depending on the degree of beam averaging.

\subsubsection{Propagation effects} \label{s:prop}

Propagation effects such as dispersion, pulse broadening (scattering) and scintillation can arise from the intervening interplanetary, interstellar and/or intergalactic media, in addition to the Earth's ionosphere. For most sources of Galactic origin, the dominant contribution comes from the interstellar medium (ISM); however, for sources at extragalactic or cosmological distances, there may also be significant contributions from the ISM of the host galaxy and/or from the intergalactic medium. All propagation effects are strongly frequency dependent and the magnitudes vary significantly with the location and direction of the source (see Cordes 2009 and Macquart 2011 for detailed discussions).

Such effects are most pronounced at low radio frequencies, and along sight lines toward the centre of the Galaxy. The dispersion delay
($\Delta t_{\rm dm}$) across the observing bandwidth $ \Delta \nu = \nu _{\rm hi} - \nu _{\rm lo} $ is given by
\begin{equation}
 \Delta t  _{\rm dm}  {\rm (ms)}  = 4.1488 \left(  \int_0^D n_{\rm e} (l) \, dl \right) \left( \nu _{\rm lo}^{-2} - \nu_{\rm hi}^{-2} \right) \approx 4.15 ~ {\rm DM} \, \left( \nu_{\rm lo}^{-2} - \nu_{\rm hi}^{-2} \right)
\end{equation}
where $\nu_{\rm hi}$ and $\nu_{\rm lo}$ are the highest and lowest frequencies (in GHz); the dispersion measure (DM) is the integrated electron density along the line of sight (units pc~cm$^{-3}$). Thus for small bandwidths ($\Delta \nu \ll \nu$), the delay scales as $\nu^{-3}$ (where $\nu$ is the centre frequency), which means significantly large delays at large DMs and at low radio frequencies. This effect must be therefore corrected prior to searching for any transients, as it can render signals potentially undetectable. As the DM is not known {\it a priori}, this requires searching over a large range in DM, e.g.\ up to $\sim 1000$~pc~cm$^{-3}$ for searches at low frequencies.

Radio waves are diffracted and refracted as they propagate through the intervening media, leading to scattering and scintillation (Rickett 1990). Likewise dispersion, interstellar contribution tends to dominate over those from others for most sources along the Galactic lines of sight. For any pulsed signals, this means a spread in arrival times as the scattered signals from a range of directions reach the telescope. This pulse broadening effect leads to asymmetric pulse shapes with the tail end of the pulse stretched out. Measured pulse broadening times ($\tau_{\rm d}$) are known to scale steeply with the frequency, $\tau_{\rm d} \propto \nu^{-(3.9\pm0.2)}$ (Bhat et al.\ 2004). While pulse broadening conserves the total flux, the smearing in time leads to smaller pulse amplitudes (i.e.\ lower peak flux densities), and hence lower signal-to-noise (S/N) in the detection. Detection of heavily scattered signals can therefore become difficult.

Scintillation effects that lead to brightening or dimming of signal strengths can potentially lead to the scenario whereby strong sources become undetectable, or may render detections of intrinsically weaker sources.  While diffractive scintillation can lead to rapid modulations of the signal amplitude in time and frequency, refractive scintillation is typically manifested as slow modulations of flux density over time scales of days or weeks
(e.g.\ Bhat, Rao \& Gupta 1999).
Regardless of their impact or influence on the detectability of signals, such propagation effects can potentially serve as important discriminators in distinguishing real astrophysical signals from those of terrestrial origin (i.e.\ RFI or instrumental effects).


\begin{table}
\caption{Taxonomy of transient detection methodologies based on number of stations and receiver design.}\label{tab:simple}
\medskip
\begin{center}
\tabcolsep 4.0pt
\begin{tabular}{lccccccc}\hline
Search & Example & No. of & No. of & Frequency & FoV & Localisation   & Sensitivity \\
Strategy & Instrument    &   stations             & pixels             & (GHz) & (${\rm deg^2}$) & radius$^{\mathrm{a}}$         & ${\rm S_{\rm sys}}$ (Jy)   \\\hline
Single-station  & GBT  & 1  & 1 & 0.35 & 0.36 & 36$^{\prime}$ & 23 \\
Single-pixel     &          &      &    &      &                         &                         & \\
               &                  &                &              &         &                &                    &          \\
Single-station  & Arecibo    &  1 & 7    & 1.4 &   0.0034 & 3.5$^{\prime}$ & 3 \\
Multi-pixel       &  Parkes &   1 & 13 &  1.4 &  0.054   & 14$^{\prime}$ & 31 \\
               &                  &                &              &         &                &                    &          \\
Multi-station    & VLBA   & 10 & 1 & 1.5  & 0.212 & 0.01$^{''}$    & 30 \\
Single-pixel     & GMRT  & 30 & 1 & 0.3  & 1.82   &  10$^{''}$ & 11 \\
               &                  &                &              &         &                &                    &          \\
Multi-station    & LOFAR  & 192 & 100 & 0.2  & 400 & 3$^{\prime}$-10$^{''}$ &  19 \\
Multi-pixel       & ASKAP   & 36   & 30  & 1.2   & 30   & 7$^{''}$                & 54 \\\hline
\end{tabular}\\[5pt]
\begin{minipage}{\textwidth}
\small Notes: Sensitivity, expressed as system equivalent flux density, $S_{\rm sys} \propto (A_{\rm eff} / T_{\rm sys})^{-1}$, is directly related to
the survey figure of merit. The field of view (FoV) for the nominal observing frequency is listed in the table. The numbers on
localisation radius and sensitivity are indicative only; $^{\rm a}$: for array instruments, this scales as ${\rm (SNR)^{-1} }$, where SNR is
the signal-to-noise ratio achieved in snap-shot imaging.
\end{minipage}
\end{center}
\end{table}

\subsubsection{Search parameter space}

The time duration of fast transients is typically quantified as the equivalent pulse width, $W_{\rm p}$. Thus, DM and $W_{\rm p}$ are the two most basic parameters required to characterize a transient signal, and therefore transient searches involve spanning a large range in both these parameters. The dispersion delays can be quite significant at low radio frequencies, e.g.\ a 1 ms pulse at DM = 10 pc~cm$^{-3}$ will be dispersed over $\approx 100$~ms in observations made across a bandwidth of 32 MHz centered at 300 MHz. At such low frequencies, the maximum DM to be searched may be quite likely limited by pulse broadening, especially for lines of sight within a few degrees of the Galactic plane.

The number of trial DMs are typically comparable to the number of spectral channels, $N_{\rm dm} \sim N_{\nu}$, though these can be more precisely determined by the spacing required to not degrade S/N significantly (this depends on the pulse width, $W_{\rm p}$, i.e.\ degradation in S/N is higher for narrow pulses). As the dispersion delays scale steeply at low frequencies ($\Delta t_{\rm dm} \sim \nu^{-3}$; see equation 3), the DM spacings tend to be smaller, requiring a large number of DMs to span a given DM range. Away from the plane, pulse broadening is not necessarily a limitation, and therefore searching out to large DMs means sensitivity to signals originating at extragalactic or cosmological distances (Cordes \& Lazio 2002; Cordes 2009).

The time duration of transient signals ($W_{\rm p}$) can vary over a wide range.  Although transient phenomena are known down to nanosecond time scales (e.g.\ giant pulses from the Crab), for searches at frequencies at or below $\sim 1$~GHz, such signals will be broadened to $\sim$ microseconds or longer due to scattering (Bhat et al.\ 2008). In practice, the shortest time scale that can be searched is essentially limited to the time resolution of the data ($W_{\rm samp}$), and any signals of much shorter intrinsic widths ($W_{\rm int} \ll W_{\rm samp}$), will essentially be instrumentally broadened to $W_{\rm samp}$. The measured pulse width $W_{\rm p}$ can be modeled as
\begin{equation}
  W_p  =   \left( W_{\rm int}^2 + W_{\rm samp}^2+ W_{\rm dm,ch}^2 + W_{\rm dm,err}^2 + W_{\rm scatt}^2 \right)^{1/2}
\end{equation}
where $W_{\rm int}$ in the intrinsic pulse width; the other terms denote the smearing due to instrument ($W_{\rm samp}$), residual dispersion ($W_{\rm dm,ch}$), error in DM ($W_{\rm dm,err}$) and pulse broadening due to scattering ($W_{\rm scatt}$). At large DMs, pulse broadening will limit the achievable time resolution, and it will be difficult to detect heavily scattered pulses owing to the S/N degradation from broadening.

\subsection{Search algorithms and detection of events} 

Most traditional search algorithms for the detection of fast transients operate on fast sampled, multi-channel (filterbank) data and hence involve performing incoherent dedispersion followed by searching for transient events using matched filtering or alternate techniques. Methodologies for both single-pixel and multi-pixel systems have been well developed and extensively applied in a number of searches (e.g.\ Cordes \& McLaughlin 2003; Bhat et al.\ 2005; Deneva et al.\ 2009; Burke-Spolaor et al.\ 2011b; Bhat et al.\ 2011). Alternate techniques based on quadratic discriminant and other statistics are also being explored (e.g.\ Thompson et al.\ 2011; Fridman 2010), though their efficacies have not yet been thoroughly tested as viable alternatives in large-scale searches. Below we describe the most basic steps in typical processing chains that are designed to detect and identify candidate events.

\subsubsection{Generation of filterbank data}

For most single-dish instruments, multi-channel time series data can be readily generated by the recording instruments, at time and frequency resolutions constrained by either data throughput requirements or the availability of processing resources. However array instruments such as VLBA and GMRT that operate on raw voltage data streams from individual elements will require significant signal processing in order to combine these data streams into multiple incoherent filterbank data streams (i.e.\ detection and channelization after incoherent or coherent combination of telescope power).
More recently, visibility-based searches have been proposed as a plausible strategy for transient searches with ASKAP, however the necessary techniques and algorithms are currently in an exploratory stage.

\subsubsection{Dedispersion over many trial DMs}

Dedispersion is the most critical and computationally intensive part in searching for fast transients. As discussed in Section~\ref{s:prop}, the dispersion delays can be quite large at low frequencies (cf.\ equation 3; $ \Delta t_{\rm dm} \propto \nu^{-3}$ for small $ \Delta \nu$) and as a result many more trial DMs are required to search even for small recording bandwidths. Most searches at frequencies $\sim 1$~GHz or lower span DM values up to $\sim 1000$~pc~cm$^{-3}$, above which the signal detection can be heavily hampered by scatter broadening. The computational efficiency achieved for the dedispersion process thus holds the key for achieving transient searching in or near real-time. Recent advances in GPU-based computing and implementation of dedispersion algorithms in such platforms provide great promise in this direction (Barsdell et al.\ 2010; Magro et al.\ 2011).

\subsubsection{Detection of transient events}

The most common method for finding transient `events' is matched filtering, where a pulse template is effectively convolved with the dedispersed time series. In principle, the template can be chosen to reflect the diversity of pulse shapes that are searched. For example,
unscattered pulse shapes can be modeled as either a single or a sum of two or more Gaussians whereas asymmetric pulse templates would be a better approximation for scattered pulses which tend to have exponential tails. In practice, most search pipelines approximate matched filtering by a hierarchical smoothing of time series by adding up to $2^n$ samples and selecting events above a set threshold after each iteration. The pulse templates are thus effectively boxcars of widths $2^n$ samples.

While relatively simple and easy to implement, matched filtering, as described above, has certain shortcomings; for example broad, strong bursts will likely be detected as an overwhelmingly large number of events. Alternatives based on time domain clustering of samples along the {\it friends-of-friends} logic can help alleviate this (e.g.\ Deneva et al.\ 2009). In this method, the dedispersed time series is processed sequentially and if an event above the threshold is found it is designated as the first of a cluster. A cluster of events is then augmented while successive samples are found to be above the threshold. The brightest sample of a cluster is recorded as the event amplitude and the total number of samples in the cluster as its width. The maximum sensitivity is however limited to the basic sampling resolution  and the maximum time duration by the size of the data block.

\subsubsection{Identification of candidate events}

\begin{figure}
\centerline{\includegraphics[width=12cm]{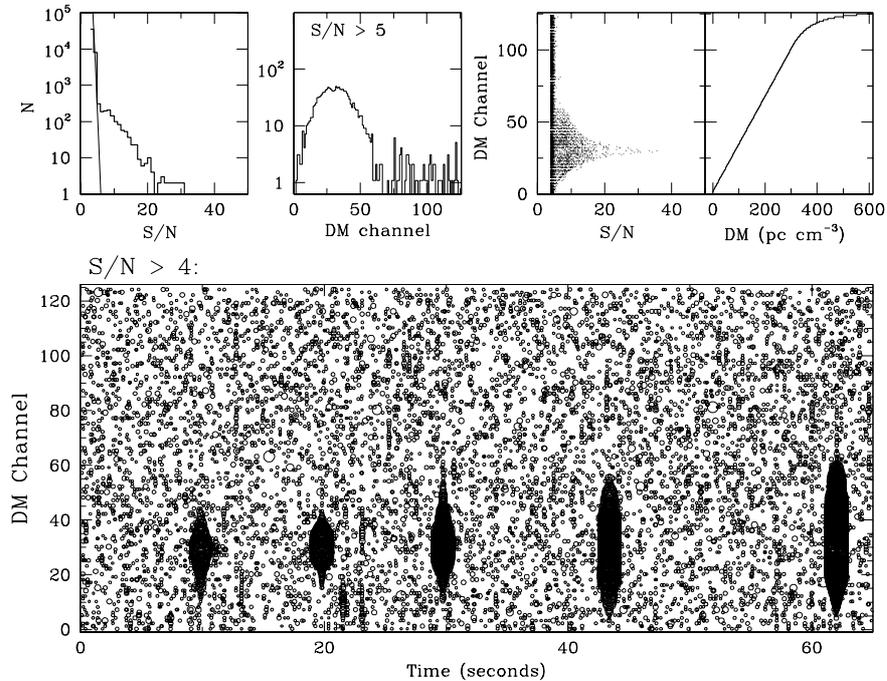}}
\vspace{-3.5cm}
\caption{Diagnostic plots from transient search methodologies for {\it single-station, single-pixel} systems (an adapted version from Bhat et al.\ 2011). The panels show detections of individual pulses from J0628+0909, an intermittent pulsar that was discovered with the Arecibo telescope (Cordes et al.\ 2006; Bhat \& Cordes 2005). The top panels (left to right) show the statistics of events above 5$\sigma$ against S/N and DM, the scatter plot of S/Ns and DMs, and the mapping from the DM channel to DM value. The bottom panel shows the summary of events (S/N $ > $ 4) in the time--DM plane. The signal strength tends to peak near the true DM of the object, $\approx 88$~ pc~cm$^{-3}$, for J0628+0909. The size of the open circle is proportional to the peak S/N.}
\label{fig:arecibo}
\end{figure}

The strategies for identifying candidate events may significantly differ depending on the instrument and the methodology adopted for searches. The ultimate goal is to identify and eliminate numerous spurious events that arise from either RFI or instrumental malfunctioning or processing artifacts. For single-pixel systems, the basic strategy will thus have to largely rely on the diagnostics from search output (see Fig.~\ref{fig:arecibo}). Any genuine signals of astronomical nature can be recognized through their distinct signatures; e.g.\ detection over a contiguous range of DMs, peaking near the true DM and with a signal strength that falls off smoothly with an increasing departure from the true DM. The width of the peak vs DM will be broader for broader pulses and the density of events on the scatter plot will depend on the number of similar events detected in the data. In contrast, RFI with narrow time structure will peak at zero DM and fall off rapidly at larger DMs unless it has intrinsic swept-frequency structure.

Multi-pixel systems such as the Arecibo L-band Feed Array (ALFA) and the Parkes multibeam receiver (PMB) allow more effective discrimination against spurious events of RFI origin through the reality checks possible via the coincidence of such search outputs across different beams (see Fig. 3). In practice however, the efficacies of such schemes depend on the optics of the telescope and the nature of the RFI environment. For instance, it is possible for RFI bursts to be detected in all, some, or none of the seven beams of ALFA, due to the complex optics of the Arecibo telescope, making implementation of such schemes rather non-trivial. For array instruments such as GMRT and VLBA, which effectively trades-off detection sensitivity to allow coincidence checks for the rejection of spurious events, the methodologies can be heavily instrument-specific. We  defer details to the relevant papers.

\subsection{Challenges and considerations in fast transient searches}

\subsubsection{Radio frequency interference}


Spurious signals arising from radio frequency interference (RFI) are a major impediment in fast transient searches.
Impulsive and powerful RFI bursts can potentially subdue, or even mask, the dispersed radio pulses from pulsars,
and often tend to manifest as spurious signals at non-zero DMs (e.g.\ Bhat et al.\ 2005). Persistent RFI signals may also adversely impact transient detection, by raising apparent detection thresholds, thus reducing sensitivities to weaker signals. With the ever increasing number of RFI sources and complex instrumentation, handling such spurious signals, in particular efficiently identifying and excising a large fraction of them, is proving to be a major challenge.  Instruments with single-pixel receiver systems can only offer limited capabilities for discrimination against spurious signals of RFI origin. However, multi-pixel receiver systems can be much more effective, as vividly demonstrated through recent discoveries of rotating radio transients (e.g.\ Deneva et al.\ 2009; Burke-Spolaor \& Bailes 2010) in searches based at Parkes and Arecibo. While most spurious signals of local origin are likely to appear in most or all pixels of the multi-beam receivers, depending on the relative offset with respect to the pixel phase centers, a real signal of astrophysical origin is likely to appear in just one beam, or at most in a few beams. This simple, albeit powerful, strategy is well exploited by both ongoing large surveys as well as targeted searches that use multi-beam receivers (e.g.\ Deneva et al.\ 2009; Burke-Spolaor et al.\ 2011b; Bhat et al.\ 2011).

Array instruments can potentially offer much higher levels of RFI rejection capabilities as their long baselines can be well exploited
for effective coincidence checks to allow identification and elimination of a large fraction of spurious events that are not common to
all array elements. This methodology is the key strategy for the transient searches being conducted with VLBA and GMRT. Further RFI
discrimination can be achieved by incorporating snap-shot imaging of candidate events as an important part of the event analysis
methodologies. As the RFI environment tends to vary significantly between the instruments, RFI detection and mitigation strategies are
also likely to be highly instrument-specific. In addition to such RFI identification and excision strategies at the post-processing
stages, significant resilience against RFI can also be developed through appropriate strategies during the data recording and
pre-processing stages. For instance, data recording systems that use multi-bit recording can be enormously advantageous due to much
higher dynamic ranges possible than traditional one or two-bit recorders used in most systems. Furthermore, with impressive advances
in computing, it may soon become feasible to consider implementing online RFI detection and excision based on schemes such as median
absolute deviation or spatial filtering (e.g.\ Roy et al.\ 2010; Kocz, Briggs \& Reynolds 2010).


\subsubsection{Data processing and computational requirements}

Searching through DM space is often the most computationally expensive part in transient search\-es, in particular at low frequencies of GMRT and LOFAR, where the number of trial DMs can be very large in order to optimally sample a given DM range. For array instruments, depending on the data rates (determined by the recording bandwidth, the number of bits per sampling and the number of array elements), the process of generating the required number of data streams (i.e.\ number of pixels across the FoV in the case of multibeaming or the number of independent data streams used for coincidence checks in the case of incoherent combination of total power), can also be a computationally intensive operation. Cordes (2009) discusses many of these considerations in detail. Since for each data stream, there are $N_{\nu}$ summations to perform for each time step $\Delta T$ (i.e.\ time resolution of the data) for every trial DM,  the total number of operations (per second) for dedispersion ($\dot{N}_{\rm dd}$) is given by
\begin{equation}
\dot{N}_{\rm dd}  \, ({\rm op\,s^{-1}}) \, \approx \, 10^{12} \, \left( { N_{\rm sa} \over 8 } \right) \left( { N_{\rm pol} \over 2 } \right) \left( { N_{\nu} \over 1024 } \right) \left( { N_{\rm dm} \over 10^3 } \right) \left( \Delta T \over 16 \, \muup s \right ) ^{-1}
\end{equation}
\noindent
where the quantities $N_{\rm sa}$ and $N_{\rm pol}$ denote the number of independent data streams and the number of polarization channels respectively; $N_{\nu}$ is the number spectral channels and $N_{\rm dm}$ is the total number of trial DMs searched.
As the dispersion delays scale steeply at low frequencies ($\Delta t_{\rm dm} \sim \nu^{-3}$), the DM spacings tend to be smaller, requiring a large number
of DMs to span a given DM range, thus translating to large computational cost for searches at low frequencies.

\subsubsection{Survey speed, search volume and optimisation of survey strategies}


The survey speed and the search volume sampled are the most critical parameters determining the figure of merit (FoM). For blind searches, the survey speed is essentially the rate of sky coverage $\dot{\Omega} \, {\rm ( deg^2 \, s^{-1} ) } = \Omega_{\rm i} / T_{\rm obs}$, where $\Omega_{\rm i}$ is the instantaneous sky coverage (i.e.\ field of view) and $T_{\rm obs}$ is the dwell time per pointing. The search volume depends on the maximum distance to which a detection
is possible ($D_{\rm max}$), which scales as $(S_{\rm pk,min})^{-1/2}$ (see equation 2). The maximum search volume is thus given by
$V_{\max} = (1/3) \Omega_{\rm s} D_{\rm max}^3$, where $\Omega_{\rm s}$ is the total sky coverage. Cordes (2009) discusses FoMs and other survey metrics for different possible survey strategies, for both fast and slow transient searches with the SKA. Another important consideration is the dependence of the search volume on temporal smearing due to scatter broadening in the ISM. For searches at low frequencies this essentially means the detection rate is a function of the Galactic position. Macquart (2011) derives a scaling for the detection rate $\Omega_{\rm i} \, S_{\rm min}^{-3/2 + \delta}$, where $S_{\rm min}$
is the minimum detectable flux density and $0 < \delta \le 3/2$ for surveys limited by interstellar scattering. These are important considerations in optimising search strategies for maximal survey yields.

While the surveys with upcoming next-generation wide-field arrays will most likely be designed and optimised based on such
considerations, ongoing surveys with large single-dish instruments and those in the near-term with existing arrays (with narrow fields
of view) provide limited or little flexibility for any such optimisation. Searching for fast transients is a secondary objective in
the high time resolution surveys underway at Parkes and Arecibo, where the survey parameters are largely optimised for maximal pulsar
discoveries (Cordes et al.\ 2006; Keith et al.\ 2010). It is encouraging however that these surveys are uncovering an increasingly
larger population of intermittent pulsars, many of which can be classified as rotating radio transients (RRATs; McLaughlin et al.\
2006; Deneva et al.\ 2009; Burke-Spolaor et al.\ 2011b; Keane \& McLaughlin 2011). The commensal nature of the surveys with VLBA and GMRT, while providing little choice for optimising the survey speed or detection rate, exemplify the most efficient ways of optimising telescope time for heavily oversubscribed instruments with limited field-of-view capabilities.


\section{Radio transient searches: current and future projects}\label{s:radio}


We now present a brief overview of the past and ongoing survey efforts aimed at detecting fast transients, as well as those proposed with upcoming next-generation instruments that will provide wide field-of-view capabilities. We will describe projects that span a range of detection methodologies, i.e.\ from those based on single-station single-pixel systems, to those that exploit the advantages of multiple pixels or multiple stations for efficient detection and rejection of spurious events that tend to arise from RFI sources and instrumental or processing artifacts.
We will highlight the scientific discoveries that have emerged as well as the technical developments that will ensue from various
exploratory projects currently underway. All these will pave the way to developing the much needed methodologies and strategies applicable for transient science in the SKA-era.

\subsection{Ongoing surveys with single-dish instruments}


As discussed in Section~\ref{s:detect}, the ability to distinguish between real signals of astrophysical origin and spurious ones that are of terrestrial or instrumental origin is paramount in transient detections. Table 1 provides a taxonomy of different possible methodologies, ranging from the simplest (i.e.\ single-station single-pixel systems) to the highly sophisticated (i.e.\ next generation radio arrays). Instruments that have no multi-beam (MB) capabilities (e.g.\ the Green Bank Telescope; GBT), or Parkes and Arecibo in the pre-MB era, are essentially single-station single-pixel systems, where reality-check capabilities against RFI-generated events are greatly limited. This is  exemplified by past searches (McLaughlin \& Cordes 2003), where the single-pixel nature of the data was a major limiting factor in conclusively establishing astrophysical origins for the detected candidates.


\begin{figure}
%
%
\centerline{\includegraphics[angle=270,width=12cm]{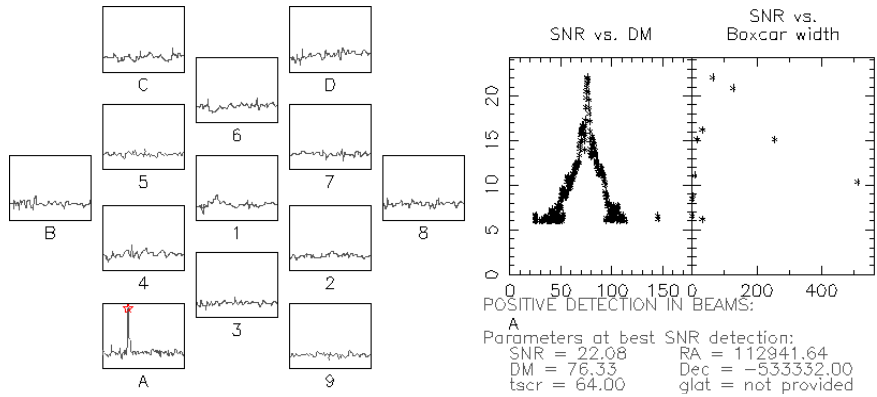}}
\caption{Example plots illustrating the transient detection methodology with {\it single-station, multi-pixel} system (an adapted version from
Burke-Spolaor et al.\ 2011b). PSR J1129$-$53 was discovered by Burke-Spolaor \& Bailes (2010) in the reprocessing of archival data from the Swinburne intermediate latitude survey (Edwards et al.\ 2001). Panels show (clockwise from top left) the dedispersed time series in all 13 beams of the Parkes multibeam receiver, the S/N versus DM (left-hand subpanel) and boxcar filter trial (right-hand subpanel) of the event. The appearance of signal in only beam A, along with clear peaks in the S/N vs.\ DM as well as S/N vs.\ the boxcar width plots, confirm the signal to be real. Spurious signals of RFI origin will typically be seen in all 13 beams.}
\label{fig:parkes}
\end{figure}

\subsubsection{High time resolution surveys at the Parkes and Arecibo radio telecopes}

The high time resolution surveys currently underway at the Parkes and Arecibo telescopes are examples of single-station multi-pixel
systems, which offer higher resilience against spurious events of RFI origin through effective anti-coincidence filtering possible by
simultaneous observations with multiple (albeit adjacent) beams (Fig.~\ref{fig:parkes}).
These surveys (i.e.\ the PALFA survey at Arecibo and the HTRU survey at Parkes) employ very similar receiver systems and instrumentation for data collection, and record data at resolutions in time and frequency that are roughly comparable. One major difference however is that the number of beams is limited to seven for Arecibo as compared to 13 for Parkes. Furthermore, with a gain that is over an order of magnitude larger, the surveys at Arecibo use much smaller dwell times than those at Parkes (i.e.\ 134--268 s for PALFA vs.\ 240--540 s for HTRU). These surveys and the re-processing of archival data from previous multi-beam surveys have already led to discoveries of a large number of intermittently emitting pulsars, the so-called RRATs (McLaughlin et al.\ 2006; Deneva et al.\ 2009; Keane et al.\ 2010; Burke-Spolaor and Bailes 2010; Burke-Spolaor et al.\ 2011b), with many more expected to emerge upon the completion of these surveys.


\subsection{Commensal surveys with array instruments}

Array instruments can potentially offer a higher resilience to RFI-generated events in transient detections. While their multi-element designs bring in a higher level of complexity in terms of the larger data rates and signal processing requirements to deal with, their long baselines and distributed nature of array elements can be exploited for effective identification and excision of spurious events that arise either from noise statistics or are of instrumental or RFI origins. Two such projects are currently underway, viz.\ the VLBA fast transients experiment and the GMRT fast transient detection project, and these will help demonstrate the efficacies of transient detection methodologies based on multi-station, single-pixel systems.

\subsubsection{The VLBA fast transient survey}

The VLBA fast radio transients experiment (V-FASTR; Wayth et al.\ 2011) uses the features and flexibility of the DiFX software correlator (Deller et al.\ 2007) to perform incoherent dedispersion on short integration ($\sim$ms) spectrometer data from different VLBA elements and search for transient events. This project is designed to run in a commensal mode, alongside routine VLBA operations. Even though the field of view is likely to  be rather small at the higher frequencies of VLBA operations (i.e.\ 1--90 GHz), the experiment offers the unique possibility to localize detected events on the sky to milli-arcsecond accuracies.


\begin{figure}
\centerline{\includegraphics[angle=0,width=\textwidth]{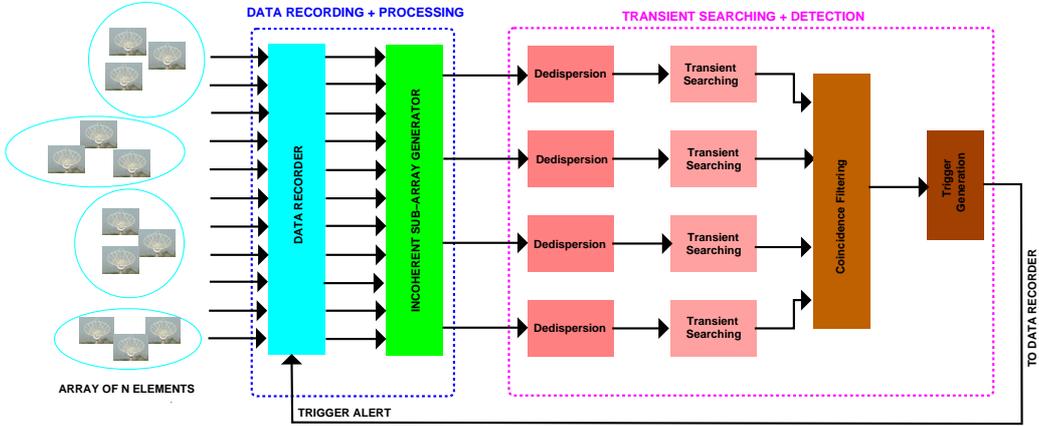}}
\caption{
Diagrammatic representation of a plausible transient detection scheme applicable for multi-element, distributed arrays such as the GMRT. Raw voltage data from each array element are captured by the data recorder and made available to a software processing pipeline. These data can be processed to form multiple  (incoherent) sub-array data streams,  which can then be processed in or near real-time for prospective transient candidates. The events from multiple parallel search pipelines  (one per each sub-array) are then assimilated and the candidate signals are identified based on a coincidence filter logic. In real-time implementation of such a scheme, this information can be used to generate trigger signals that will alert the data recording system to store the relevant raw data segments for further detailed processing and scrutiny.}
\label{fig:pipeline}
\end{figure}

\subsubsection{The GMRT transient survey}

\begin{figure}
\centerline{\includegraphics[angle=270,width=\textwidth]{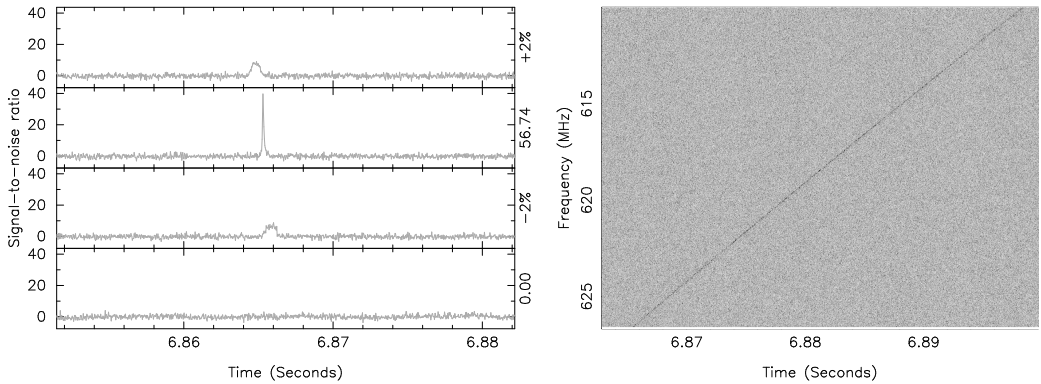}}
\caption{Example plots from the GMRT transient detection pipeline -- an illustration of {\it multi-station, single-pixel} search methodology. The array was configured to form seven independent sub-arrays, each comprising just a single antenna, thus providing a powerful coincidence filter against spurious events of RFI origin. Data were collected by emulating a `survey mode', by making the array to scan the sky region around the Crab pulsar at 0.5 ${\rm deg~min^{-1}}$. A bright giant pulse was thus detected as a `transient' when the pulsar was within the telescope beam (half power beam width $\sim 0.5^{\circ}$). The pulse is very narrow ($\approx 50$ $\muup$s), i.e.\ approximately twice the sampling resolution (30 $\muup$s) and so seen as a thin strip on the waterfall plot. The signal peaks at DM = 56.74 pc~cm$^{-3}$, with a sharp decline in S/N even at small departures from the true DM; for instance, ${\rm \Delta DM / DM \approx 0.02}$ results in a S/N loss of almost a factor of four, exemplifying the need for very short DM spacings and high time and frequency resolutions for transient searches at low frequencies.}
\label{fig:gmrt}
\end{figure}

The transient exploration project currently underway at the GMRT is yet another effort aimed at demonstrating the efficacies of
multi-station single-pixel methodologies for transient detections. This low-frequency array of $30 \times 45$-m paraboloidal dishes,
operating at frequency bands from 0.15 to 1.5 GHz, has an effective collecting area ($A_{\rm eff}$) that is $\sim 3$\% of the
SKA, and thus has the potential to enable sensitive transient searches. The array offers several unique design features that can be
well exploited for transient exploration. In particular, its moderate number of elements, long baselines (up to $\sim 25$~km) and
flexible sub-array capabilities make it a perfect platform for developing and demonstrating novel methodologies and strategies
applicable for SKA-era transient science. The advent of a new software backend (Roy et al.\ 2010), in particular its ability to record
raw voltage data from individual array elements and make them available for software-based processing systems, opens up excellent
avenues for such demonstrator experiments (Figs.~\ref{fig:pipeline} and \ref{fig:gmrt}). Details on the development of the transient detection pipeline and pilot surveys are described in a forthcoming paper (Bhat et al.\ 2011, in preparation).


\subsection{Surveys with next-generation (wide FoV) instruments}

Large-element, distributed arrays are the future of radio astronomy and most next-generation radio arrays are being designed to offer wide field-of-view capabilities -- an important requirement for conducting efficient transient explorations. While low-frequency arrays such as the MWA and LOFAR will have fields of view of the order of hundreds of ${\rm deg^2}$, ASKAP with its focal plane arrays will offer a field of view $\sim 30$~deg$^2$. Multi-beaming is an important consideration in the LOFAR design as well as for the ASKAP and MeerKAT systems, with the possibility of forming of hundreds of tied-array beams with the LOFAR stations. These next-generation arrays will thus offer excellent avenues for efficient transient exploration and will benefit from methodologies that exploit their multi-station, multi-pixel advantages. As previously discussed by several authors (Cordes 2009; Hessels et al.\ 2009; Macquart et al.\ 2010), it is useful to define a figure of merit (FoM) that distills the various observational requirements for transient surveys.
\begin{equation}
 {\rm FoM} \propto \left( { A_{\rm eff} \over T_{\rm sys} } \right)^2 \, { \Omega_{\rm i} \over \Delta \Omega } \, { T_{\rm obs} \over \Delta T}
\end{equation}
where $\Omega_{\rm i}$ is the instantaneous field of view and $\Delta \Omega$ is the localization radius; $T_{\rm obs}$ and $\Delta T$ are the dwell time
and time resolution of the data, respectively. Thus, to effectively probe transients over a wide range of source parameter space, FoM
needs to be maximized, which in turn means a large effective collecting area, $A_{\rm eff}$ (i.e.\ raw sensitivity) and instantaneous field of view (FoV) in addition to longer dwell times, while also maintaining high resolutions in both time and on sky.

\subsubsection{The LOFAR transient survey}

Pulsar and fast-transient searches are a key science project for LOFAR. A variety of observing modes are possible given the flexible digital designs of the hardware, and thus FoM may vary significantly depending on the incoherent vs.\ coherent combination of telescope power. While the possibility of forming hundreds of tied-array beams will potentially enable high-instantaneous-sensitivity all-sky surveys, the survey speed would be significantly limited due to smaller fields of view in such modes. The use of multiple incoherent beams would imply shallower surveys in terms of sensitivity; however, with large fields of view, much longer dwell times will be feasible. Details of various possible observing modes and the science case for transient exploration are described in recent LOFAR papers (Fender et al.\ 2008; Hessels et al.\ 2009; Stappers et al.\ 2011).



\subsubsection{The ASKAP CRAFT survey}


The Australian SKA Pathfinder (ASKAP) -- an upcoming array of $36 \times 12$-m dishes in Western Australia -- is currently the only wide field-of-view instrument that is designed to operate in the mid-range frequencies (i.e.\ 0.7--1.8 GHz). With its unprecedented combination of sensitivity and a $\sim 30$~deg$^2$ FoV, ASKAP will be a powerful instrument for transient exploration in the southern sky. Among the key science projects with ASKAP is the Commensal Real-time ASKAP Fast Transient (CRAFT) survey (Macquart et al.\ 2010) -- a project that is being designed to run in a purely commensal mode and will search for transients on time scales of milliseconds to seconds. In addition to traditional searches that will be  possible through incoherent or coherent combination of telescope power, the CRAFT project also proposes to conduct coherent searches using high resolution visibility data ($\approx 1$~ms, $\approx 1$~MHz) obtained from the correlator. The ASKAP project is currently in its advanced engineering prototyping stage and the array is expected to become fully operational by 2014.

\subsection{SKA as a Radio Synoptic Survey Telescope}



With sensitivity, imaging resolution and survey speed that is one to two orders of magnitude larger than almost all currently operational radio telescopes, the SKA will indeed provide an unprecedented opportunity for conducting in-depth explorations of the transient radio sky.  Cordes (2009) presents a convincing and compelling case to design the SKA as a radio synoptic survey telescope (RSST). Even though much of the discussion and the schemes outlined therein are mainly within the context of the Large-N-Small-D (LNSD) concept that is being advocated for the SKA, i.e.\ science case for mid-range frequencies (e.g.\ 0.3 to 3 GHz), the basic idea is to maximally exploit the combination of high sensitivity, wide fields of view and flexible signal processing systems for a wide variety of transient and pulsar science goals (and even the detection of ETI signals). In addition to multi-station and multi-beaming advantages of pathfinder-class instruments such as LOFAR and ASKAP, the SKA will offer additional avenues through the possibility of pixelization (multibeaming) across the field of view (i.e.\ phased array sensitivity across the full field of view), very long baselines extending up to thousands of kilometres and flexible sub-arrays that allow optimal trade-off of sensitivity vs.\ the survey speed. The example synoptic cycle described in this SKA memo underscores the need for field-of-view expansion through use of multiple-pixel receivers in order to conduct a large-scale survey in a reasonable amount of time, and also outlines the associated calibration and processing requirements.

\section{Concluding remarks} \label{s:conc}

The transient radio sky is largely an uncharted territory with tremendous untapped potential for exciting discoveries and innovative science. The two key requirements for enabling comprehensive explorations of the transient radio sky are
wide-field capability and high instantaneous sensitivity, which most upcoming next-generation radio facilities will offer.
Such instruments can potentially revolutionize our knowledge of the transient radio sky in the coming decades.

Signal processing requirements can be quite phenomenal in searches for short-duration (fast) radio transients, in particular at low frequencies (i.e.\ 0.1--1 GHz) where propagation effects are severe and RFI-related challenges are numerous. Fortuitously, with impressive advances in super computing of recent years, this is becoming less of a challenge. Nonetheless, it is important to devise effective data archival strategies and avenues for long term data curation and mining in order to ensure long term science pay-offs. The development of {\it cyberinfrastructure}, such as for example, the upcoming Pawsey HPC Centre for SKA Science and Computing in Perth, Australia, can thus be expected to play an indispensable role in this area of research.

As the design and construction of next-generation instruments move forward, it is imperative to sustain the scientific and technical development through the effective use of existing instruments. While their limited fields-of-view and available telescope time pose major constraints, the technological advances can be exploited for devising suitable observing and processing strategies to ensure rapid turn-around in science outcomes. The high time resolution searches with Arecibo and Parkes that are uncovering an increasingly large population of intermittently emitting pulsars are a vivid demonstration of such success strategies. In the near-term, instruments such as VLBA and GMRT will serve as excellent test beds for developing and demonstrating useful methodologies and strategies for transient detections with multi-element, distributed arrays. In the longer term, with their large survey speeds, instruments such as LOFAR and ASKAP are promising to initiate routine all-sky surveys. All such efforts can be expected to culminate into valuable inputs and lessons as the SKA science case is  developed and matured over the next decade.

\section*{Acknowledgements}

I would like to thank the editors, D. J. Saikia and D. A. Green, for inviting me to write this article. I also like to take this opportunity to express my gratitude to Jim Cordes for inspiring me into this exciting field of radio transients, and for generously providing the phase space plot for inclusion in this article. I thank Matthew Bailes for his continued support and encouragement toward various projects over the past years, and Peter Cox for assistance with preparation of some figures used in this article. This work is supported by the Australian Government under the Australia-India Strategic Research Fund grant ST020071. The Centre for All-sky Astrophysics is an Australian Research Council Centre for Excellence, funded by CE11E0090. The GMRT is operated by the National Centre for Radio Astrophysics (NCRA) of the Tata Institute of Fundamental Research (TIFR), India. Finally, I like to acknowledge my collaborators from Swinburne, NCRA and other institutions in the GMRT transient project.


\label{lastpage}
\end{document}